\documentclass{svjour2}                    % onecolumn
\smartqed  % flush right qed marks, e.g. at end of proof
\usepackage{graphicx}
\begin{document}

\title{Spin down of superfluid-filled vessels: theory versus experiment }


\author{C. A. van Eysden          \and
        A. Melatos %etc.
}


\institute{C. A. van Eysden \at
              School of Physics, University of Melbourne, Parkville, VIC, 3010, Australia \\
              \email{ave@unimelb.edu.au}           %  \\
           \and
           A. Melatos \at
              School of Physics, University of Melbourne, Parkville, VIC, 3010, Australia \\
              \email{amelatos@unimelb.edu.au}  
}

\date{Received: date / Accepted: date}

\maketitle

\begin{abstract}
The spin up of helium II is studied by calculating the spin-down recovery of a superfluid-filled container after an impulsive acceleration and comparing with experiments.
The calculation takes advantage of a recently published analytic solution for the spin up of a Hall-Vinen-Bekharevich-Khalatnikov superfluid that treats the back-reaction torque exerted by the viscous component self-consistently in arbitrary geometry for the first time.
Excellent agreement at the $0.5\%$ level is obtained for experiments at $T=1.57\,{\rm K}$, after correcting for the non-uniform rotation in the initial state, confirming that vortex tension and pinning (which are omitted from the theory) play a minimal role under certain conditions (small Rossby number, smooth walls).
The dependence of the spin-down time on temperature and the mass fraction of the viscous component are also investigated.
Closer to the lambda point, the predicted onset of turbulence invalidates the linear Ekman theory.

\end{abstract}

\section{Introduction}
\label{tsasec1}

%Good agreement between theory and experiment in helium II is difficult to obtain.
The spin up of helium II in closed vessels has been studied experimentally \cite{hal57,wal58,hal60,pel60,rep60,tsa72,tsa73,tsa80} and theoretically \cite{cam82,ada85,per90,rei93}, but good agreement has proved elusive.
The problem poses a significant challenge over classical Navier-Stokes spin up because of the additional complications arising from the two-component hydrodynamics (coupled by mutual friction) in a superfluid and quantum mechanical vortex pinning and nucleation.
Early studies of helium II spin up measured the torque required to accelerate a fluid-filled container, while monitoring the fluid response with a Rayleigh disk device suspended in the flow \cite{hal57,wal58,hal60,pel60}.
Asymmetric acceleration and retardation and strong hysteresis were observed, effects which are enhanced in rough-walled containers, suggesting that vortex pinning and nucleation play a crucial role.
% When a container is accelerated from rest, several minutes elapse before the fluid responds; in contrast, the fluid response occurs immediately, when the rotation is stopped \cite{pel60,rep60}.

The first comprehensive experimental study of the spin up of helium II was conducted by Tsakadze and Tsakadze \cite{tsa72,tsa73,tsa80}, who monitored the hydrodynamic relaxation of superfluid-filled containers after an impulsive increase in angular velocity.
% The experiments were designed to shed light on terrestrial helium II as well as the role of superfluidity in pulsars.
Spherical and cylindrical containers made of plexiglas or duralumin were used; an additional coating of plexiglas powder was applied to study strong vortex pinning.
For smooth-walled containers, a clear decrease in the spin-up time with increasing temperature is observed.
The spin-up time is also a function of initial and final angular velocity, container size, and density fraction of the viscous component.
In rough-walled containers, a significantly shorter relaxation time is observed, which does not depend on the temperature or velocity jump.
The above investigations have been repeated in spherical Couette geometry \cite{tsa80}.

No complete theoretical explanation of the Tsakadze results has been published to date.  
The lack of a theory is felt most keenly in the field of neutron star astrophysics, where the Tsakadze experiments serve as a terrestrial facsimile of the glitch phenomenon.
The origin of glitches is a fundamental 40-year-old puzzle \cite{van10} which relates to the hypothesis of nuclear superfluidity \cite{van10}.
Pioneering attempts to theoretically understand the nature of spin up in helium II focused on the interaction between quantized vortices in the superfluid and the boundaries \cite{cam82,ada85}.
A rigorous analytical treatment was provided by Reisenegger \cite{rei93}, who solved the vortex-fluid equations of motion of Chandler and Baym \cite{cha86} between two infinite, rotating parallel plates accelerating at a slow and constant rate.
In smooth-walled containers, two spin-up mechanisms operate; classic Ekman pumping in the viscous component near the superfluid transition temperature, and Ekman-like circulation in the inviscid component near absolute zero.
In the latter case, mutual friction between the vortex lines and the Ekman layer produce the secondary flow, which transports the vortex lines inwards, in agreement with Ref.\cite{ada85}.
The analysis agrees qualitatively with the experiments of Tsakadze, but detailed quantitative agreement remains elusive because the theory does not treat the feedback torque on the container self-consistently in the correct geometry, nor is it valid in the non-linear regime.

In this paper, we calculate the spin-down recovery of a superfluid-filled container self-consistently and compare our findings with the Tsakadze experiments \cite{tsa80}.
The calculation, whose details can be found in Ref. \cite{van11a,van11b}, solves the Hall-Vinen-Bekharevich-Khalatnikov (HVBK) two-fluid equations with Hall-Vinen mutual friction in arbitrary geometry.
The coupled acceleration of the fluid and deceleration of the container are followed self-consistently by continually updating the fluid boundary conditions in response to the viscous back-reaction torque.
A summary of the analytic model is given in \S\ref{tsasec2}.  
In \S\ref{tsasec3}, the angular velocity versus time, and the dependence of the spin-down time on temperature, are compared with measurements from Ref. \cite{tsa80}.
Excellent agreement is obtained after correcting for the non-uniform rotation in the initial state.

\section{Spin-down theory of a two-component superfluid and its container}
\label{tsasec2}

Consider a cylindrical or spherical container rotating with angular velocity $\Omega_s$ about the $z$ axis (unit vector ${\bf k}$).
We assume that the superfluid corotates uniformly with the container initially ($t<0$), as in all such analyses; it turns out that experimental data force us to revisit this assumption, as discussed in \S\ref{tsasec3}.
At $t=0$, the container is spun up impulsively to $\Omega_s +\delta\Omega$ and then left to respond to the viscous and frictional torques exerted by the fluid and the apparatus.
The spin up is linear, with Rossby number $\epsilon=\delta\Omega/\Omega_s\ll1$.
The angular velocity of the container for $t\geq0$ is denoted $\Omega(t)=\Omega_s+\delta\Omega f(t)$, where $f(t)$ is the dimensionless angular velocity perturbation of the container.
At time $t=0$, when the container is released after being accelerated impulsively, the initial velocities of the viscous and inviscid components are zero in the rotating frame, and $f(0)$ equals unity.

In the model, it is assumed that the container initially rotates with Ekman number $E\ll1$.
The Ekman number is defined as
\begin{equation}
 E=\frac{\nu_n}{L^2 \Omega_s} \label{tsa201}
\end{equation}
where $\nu_n$ is the kinematic viscosity the viscous component, and $L$ is the appropriate length scale (the radius $a$ for a sphere or the half-height $h$ for a cylinder).
When $E$ is small, the interior fluid spins up via Ekman pumping.
It is customary to use the dimensionless variable $\tau$, which is the usual time-coordinate normalized by the Ekman time $E^{-1/2}\Omega_s^{-1}$ \cite{gre63,van11a}.

\subsection{Governing equations} \label{tsasec2a}

The response of the superfluid inside the container is described by the incompressible HVBK equations, which model the superfluid in terms of viscous and inviscid components coupled by a mutual friction force.
Linearized and in the rotating frame, the HVBK equations take the dimensionless form \cite{per08,van11a}
\begin{eqnarray} 
E^{1/2}\frac{\partial {\bf v}_n}{\partial \tau}+ 2 {\bf {\bf k}} \times {\bf v}_n&=&-\nabla p_n +E \nabla^2 {\bf v}_n+\rho_s {\bf F}\, , \label{goeq1} \\
E^{1/2}\frac{\partial {\bf v}_s}{\partial \tau}+ 2 {\bf k} \times {\bf v}_s&=&-\nabla p_s-\rho_n {\bf F}\, , \label{goeq2} \\
\nabla\cdot{\bf v}_n&=&0\, , \label{goeq3} \\
\nabla\cdot{\bf v}_s&=&0\, . \label{goeq4}
\end{eqnarray}
where ${\bf v}_{n,s}$ and $p_{n,s}$ are the bulk velocities and pressures of the viscous $(n)$ and inviscid $(s)$ components.
We assume laminar flow, where the linearized mutual friction force has the Hall-Vinen form \cite{hal56b}
\begin{equation} \label{goeq5}
 {\bf F}=B  {\bf k}\times\left[{\bf k}\times\left({\bf v}_n-{\bf v}_s\right)\right] +B' {\bf k}\times\left({\bf v}_n-{\bf v}_s\right) \,.\label{goeq5a}
\end{equation}
and $B$, $B'$ are the mutual friction coefficients.
The velocity and pressure scales are chosen to be $L\delta\Omega_s$ and $\rho\Omega_s L^2 \delta\Omega_s$ respectively, where $\rho$ is the total mass density for the superfluid.
The mass densities of the viscous and inviscid components, $\rho_n$ and $\rho_s$, are scaled so that
\begin{equation}
 \rho_n+\rho_s=1\,.
\end{equation}

To keep the problem analytically tractable, the walls of the container are assumed to be smooth, and vortex tension is neglected, an assumption which is justified in detail below.
The viscous component obeys the usual no-slip and no-penetration boundary conditions with respect to the tangential and normal velocity respectively.
For the inviscid component, no penetration is sufficient.

\subsection{Ekman pumping} \label{tsasec2b}

In a two-component superfluid, Ekman pumping proceeds as follows.
In the viscous component, the Coriolis force drives a secondary flow in two layers of thickness $E^{1/2} L$ along the upper and lower boundaries of the container.
To satisfy continuity requirements, the outflow in the boundary layer draws in fluid from the interior, spins it up, and cycles it back into the interior on the Ekman time-scale $E^{-1/2}\Omega^{-1}_s$.
The inviscid component is spun up by the viscous component via the mutual friction force.
In helium II the mutual friction force is strong [$B$ and $B'$ are $O(1)$], and the inviscid component is dragged along with the viscous component as it undergoes Ekman pumping.
The azimuthal velocities of both fluid components are ``locked together'' over the Ekman time-scale, behaving essentially as a single fluid \cite{van11a}.
The dimensionless quantity that controls the strength of the coupling is $2 B E^{-1/2}/(2-B')$ (see below).

The general analytic solution to (\ref{goeq1})--(\ref{goeq4}) has been derived recently \cite{van11a}, generalizing the boundary layer analysis of Greenspan \cite{gre63}.
The solution is most neatly expressed in terms of the total mass flux, defined as
\begin{equation}
 {\bf v}=\rho_n {\bf v}_n +\rho_s {\bf v}_s\,.
\end{equation}
The azimuthal component of the total mass flux is given by
\begin{eqnarray}
v_{\phi}(r,\tau)&=&\frac{r \omega_+(r) \omega_-(r)}{\beta \left[\omega_+(r)-\omega_-(r)\right]} \int_0^\tau\,d\tau' f(\tau') \nonumber\\
 && \times\left\{  \left[\omega_+(r)+\beta\right] e^{\omega_+(r)\left(\tau-\tau'\right)}  -\left[\omega_-(r)+\beta \right]e^{\omega_-(r)\left(\tau-\tau'\right)}\right\} \nonumber \\
 && -\frac{r \Omega_{n0} \omega_+(r) \omega_-(r) }{\beta\left[\omega_+(r)-\omega_-(r)\right]} \left[e^{\omega_+(r) \tau}- e^{\omega_-(r) \tau}\right] \nonumber \\
 && -\frac{r \Omega_0}{\omega_+(r)-\omega_-(r)}\left[\omega_-(r)e^{\omega_+(r) \tau}-\omega_+(r) e^{\omega_-(r) \tau}\right] \,,   \label{goeqa01}
\end{eqnarray}
where we work in cylindrical coordinates denoted $(r,\phi,z)$ throughout this paper.
In (\ref{goeqa01}), we make the auxiliary definitions
\begin{eqnarray}
 \omega_{\pm}&=&-\frac{1}{2}\left[\beta+\frac{I(r)}{h(r)}\right]\pm \left\{\frac{1}{4}\left[\beta+\frac{I(r)}{h(r)}\right]^2-\frac{\beta  J(r)}{h(r)} \right\}^{1/2} \, ,  \label{tsa206} \\
 \beta&=&\frac{2 B E^{-1/2}}{2-B'} \, , \label{tsa207}\\
 I(r)&=&\frac{1}{\lambda_{-}(r)\left[\lambda_{-}^2(r)+\lambda_{+}^2(r)\right]  } \nonumber \\
 &&\times \left\{\left[\frac{1-H^4(r)}{1+H^4(r)}\right]^2\left[\lambda_{+}^2(r)-\lambda_{-}^2(r)\right]^2+4\lambda_{+}^2(r)\lambda_{-}^2(r)\right\}^{1/2}\, , \label{tsa208}\\
 J(r)&=&\frac{\rho_n H^2(r)}{2 \rho \lambda_{-}(r) \left[1+H^4(r)\right]} \nonumber \\
 &&\times \left\{ \left[\lambda_{-}^2(r)+\lambda_{+}^2(r)\right]\left[1-H^4(r)\right]+4 \lambda_{-}^2(r) H^4(r)\right\} \, , \label{tsa209} \\
 \lambda_{\pm}(r)&=&\frac{\rho^{1/2}}{H(r)}\left\{ \left[\frac{\left(B'-2\right)^2+B^2}{\left(\rho_n B'-2 \rho \right)^2+\left(\rho_n B\right)^2}\right]^{1/2} \right. \nonumber \\
 && \left.  \mp \frac{2 \rho_s   B \left[1+H^4(r)\right] }{H^2(r)\left[\left(\rho_n B'-2\rho\right)^2+\left(\rho_n B\right)^2\right] }\right\}^{1/2} \,, \label{tsa210} 
\end{eqnarray}
Equations (\ref{goeqa01})--(\ref{tsa210}) are valid for a container of arbitrary shape.
The fluid is contained within the region $z=\pm h(r)$.
For a sphere, we have $h(r)=H(r)^{-2}=(1-r^2)^{1/2}$.
For a cylinder, we have $h(r)=H(r)=1$ \cite{van11b}.
For axisymmetric containers such as spheres and cylinders, the solution is independent of the azimuthal angle $\phi$.

The solution (\ref{goeqa01}) is columnar (i.e., independant of $z$) as a result of the Taylor-Proudman theorem.  
In Ref. \cite{van11a} it is demonstrated that for a mutual friction force of the form (\ref{goeq5a}), each of the fluid components is columnar, extending the Taylor-Proudman theorem to the two-fluid problem.

\subsection{Back-reaction torque on container} \label{tsasec2c}

Equations (\ref{goeqa01})--(\ref{tsa210}) give the solution to (\ref{goeq1})--(\ref{goeq4}) in terms of an arbitrary $f(\tau)$.
The motion of the container $f(\tau)$ is determined by the viscous back reaction torque applied by the superfluid.
Integrating the viscous stress tensor on the boundary $z=\pm h(r)$, torque balance yields \cite{van11b}
\begin{equation} 
 \frac{d f(\tau)}{d \tau}= -\frac{4\pi K}{I_f}\int_{0}^{R}d r\,r^2 h(r)\frac{\partial v_{\phi}(r,\tau)}{\partial \tau}  + T_0\, , \label{goeq13}
\end{equation}
where we make the definition
\begin{equation}
 K=\frac{I_f}{I_c}\,,
\end{equation}
and $R$ denotes the dimensionless, equatorial radius of the vessel ($a/h$ for a cylinder of height $2h$ and radius $a$; unity for a sphere).
The constant frictional torque in the apparatus is denoted by $T_0$ in dimensionless units and is negative \cite{van11b}.
$I_f$ and $I_c$ denote the dimensionless nominal moments of inertia of the fluid (rotating as if it were a rigid body of uniform density $\rho$) and the container respectively, scaled to $\rho L^5$.
For a sphere, we have $I_f=8\pi/15$; for a cylinder, we have  $I_f=\pi R$.

Equations (\ref{goeqa01}) and (\ref{goeq13}) constitute a closed pair of integro-differential equations for the unknowns $v_\phi(r,\tau)$ and $f(\tau)$.
Substituting (\ref{tsa210}) into (\ref{goeq13}), the integral equation of motion for the container is \cite{van11b}
\begin{equation}
 f\left(\tau\right)=-K \int_0^\tau {\rm d\tau'} \, \left[ \dot{g}^A(\tau-\tau')+\dot{g}^B(\tau-\tau')\right] f(\tau')+ T_0 \tau  + 1 \,, \label{tsa203}
\end{equation}
with
\begin{eqnarray}
 g^A(\tau)&=& -\frac{4 \pi }{I_f}\int_0^R \frac{{\rm dr}\,r^3 J(r) \left[  e^{\omega_+(r)\tau}-e^{\omega_-(r)\tau} \right]}{\omega_+(r)-\omega_-(r)} \,,  \label{tsa204} \\
 g^B(\tau)&=& -\frac{4 \pi }{I_f}\int_0^R \frac{{\rm dr}\,r^3 \left\{ \omega_-(r) \left[e^{\omega_+(r)\tau}-1\right]-\omega_+(r)\left[ e^{\omega_-(r)\tau}-1\right] \right\} }{\omega_+(r)-\omega_-(r)}  \,. \label{tsa205}
\end{eqnarray}
An overdot symbolizes a derivative with respect to $\tau$.

Equation (\ref{tsa203}) captures self-consistently the back-reaction of the superfluid on the container and vice versa, e.g. by continuously varying the boundary conditions on the superfluid as $\Omega(\tau)$ evolves.
It is straightforward to solve (\ref{tsa203}) numerically, by guessing an initial trial function for $f(\tau)$, substituting it into the right-hand side of (\ref{tsa203}), updating $f(\tau)$ via underrelaxation \cite{press}, and iterating.
This procedure is described in Ref. \cite{van11a}.

\subsection{Vortex tension} \label{tsasec2d}

In deriving  (\ref{tsa203}), the vortex tension has been neglected.
The reasons for this are discussed in detail in Ref. \cite{van11a}, and we re-iterate them here.

In the solution of Reisenegger \cite{rei93} for slow Ekman pumping between slowly accelerating plates, the tension coefficient $\nu_s$ {\it only} appears in the inviscid velocity component in the boundary layer
\footnote{In fact, $\nu_s$ only enters in the combination $\nu_s/\nu_n$, which is the ratio of the tension force to the viscous force.
Although the length-scale of the system may enter the force ratio in principle, it does not do so in practice.
Chandler and Baym \cite{cha86} wrote the linearized tension force as $2 \Omega \nu_s (\partial/\partial z)^2 {\bf \xi}$ (where ${\bf \xi}$ is the vortex displacement vector), which is proportional to inverse length squared and hence scales identically with the viscous force.
Alternatively, the Hall-Vinen form of the tension is given by $\nu_s \omega_s \times (\nabla \times \hat{\bf \omega})$, which also scales in inverse proportion to length squared.}.
The qualitative features of the slow-acceleration boundary problem are expected to carry over to the impulsive spin-up problem, just like they do in the classical case \cite{bon48}.
Hence, $\nu_s$ does not appear in the solution for the interior flow or the viscous velocity component in the boundary layer.
Nor, consequently, does it appear in the spin-up time.
[In section 2 of Ref. \cite{van11a}, the results of Reisenegger \cite{rei93} are recovered despite assuming $\nu_s=0$.]
Thus, vortex tension is not neglected on the basis of scaling but by realizing that, during the spin up of smooth-walled containers, the tension $\nu_s$ appears {\it only} in the inviscid component boundary layer and therefore does not affect the gross dynamics (e.g. spin-up time)
\footnote{The ratio of the tension force to the mutual friction force is also small (of order $\nu_s/L^2 \Omega \sim E$ for the linear problem in helium II).}.

Also, because the torque is calculated from the viscous-component boundary layer (which is independent of $\nu_s$), the feedback calculated in \S\ref{tsasec2c} and Ref. \cite{van11b} is also independent of $\nu_s$.
In \S\ref{tsasec3}, we find excellent agreement between theory and experiment, supporting the hypothesis that vortex tension does not play a significant role in the linear spin-up of superfluids in smooth-walled containers at temperatures below $1.6\,{\rm K}$.

Nevertheless, it is interesting to evaluate the characteristic strength of the tension force relative to other forces.  
For a neutron superfluid we have $\nu_s\sim10^{-7}\,{\rm m^2\,s^{-1}}$, therefore, $\nu_s/\nu_n\sim10^{-12}$.
Clearly, for astrophysical systems like neutron stars, we find that the vortex tension is negligible.
However, for laboratory experiments with helium II, we find $0.5<\nu_s/\nu_n<6.1$ for $1.3\,{\rm K}<T<2.1\,{\rm K}$.
Hence, if it were not for the form of the tension force discussed in the previous paragraph, it would have to be included.
An interesting avenue for future work is to repeat the calculation in Ref. \cite{van11a} keeping the vortex tension, to clarify its role in impulsive spin up.

Because vortex lines must emerge perpendicular to the surface, there is a region close to the surface where vortex lines are not vertical in non-parallel-plate geometries.
The thickness of this region is governed by the vortex line tension.
By analogy with the classical viscous penetration depth $\delta_n=(\nu_n/\Omega)^{1/2}$, we define a vortex tension penetration depth $\delta_s$, which in helium II takes the value
\begin{equation}
 \delta_s = 2.4\times10^{-2} \left( \frac{\nu_s}{1.1\times10^{-7}\, {\rm Pa\,s}} \right)^{1/2} \left( \frac{\Omega}{1\,{\rm rad\,s^{-1}}} \right)^{-1/2}\,{\rm cm}\,.
\end{equation}
in agreement with Ref. \cite{rei93}.
Because $\nu_s$ and $\nu_n$ are comparable in helium II, this region is of approximately the same thickness as the Ekman layer.
% This compares with the classical viscous penetration depth $\delta_n=\sqrt{\nu/\Omega}$, which spans the range from $<\delta_n<1.6\times10^{-2} \,{\rm cm}$ for $1.2<T<1.6\,{\rm K}$.

\section{Helium II experiments} 
\label{tsasec3}

\subsection{Vessel rotation curve} 
\label{tsasec3a}

In the Tsakadze experiments, spherical and cylindrical containers are studied under a variety of initial conditions.
Only those experiments in which the container is initially rotating are examined in this paper, to ensure that $E$ and $\epsilon$ are small and the analytic solution in \S\ref{tsasec2} is valid.
For the helium II experiments of the Tsakadze group \cite{tsa80}, we find
\begin{equation}
 E =3.5\times10^{-5} \left( \frac{\nu_n}{5.5\times10^{-8}\, {\rm Pa\,s}} \right) \left( \frac{\Omega}{1\,{\rm rad\,s^{-1}}} \right)^{-1} \left(\frac{L}{4\,{\rm cm}}\right)^{-2}\,,
\end{equation} 
at $T=1.6\,{\rm K}$.  For $1.2\,{\rm K}\leq T\leq 1.6\,{\rm K}$, one measures $1.197\times10^{-8}\leq \nu_n\leq 8.997\times10^{-9}$ using vibrating wire viscometers \cite{tou63,goo73} (see also Donnelly's website\footnote{http://darkwing.uoregon.edu/~rjd/vapor13.htm}).
Therefore the requirement $E\ll1$ is valid for the experiments we consider here.

\begin{figure} [h!]
  \includegraphics[width=0.45\textwidth]{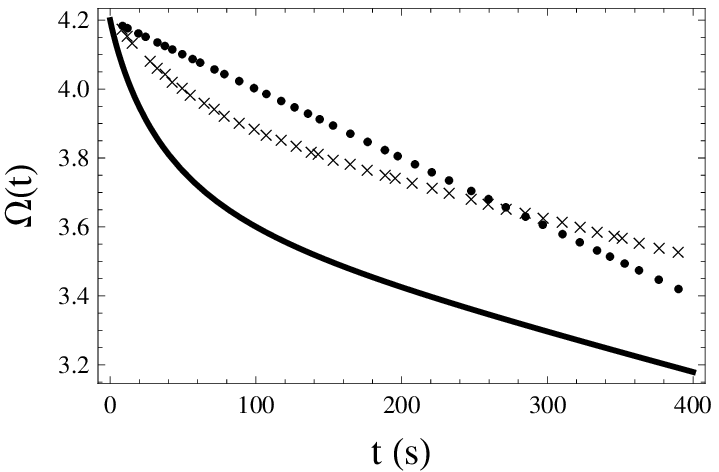} 
  \includegraphics[width=0.45\textwidth]{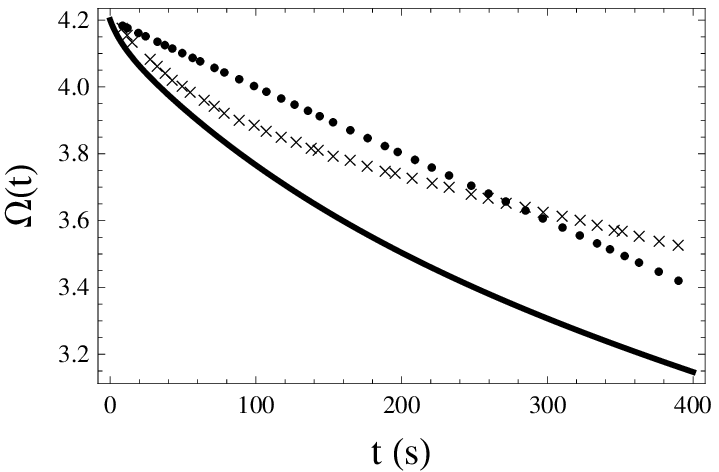}
\caption{Angular velocity $\Omega(t)$ (in ${\rm rad\,s^{-1}}$) of a spherical container versus time (in ${\rm s}$) following an impulsive acceleration from $\Omega_s=3.0\,{\rm rad\,s^{-1}}$ to $\Omega_s+\delta\Omega=4.2\,{\rm rad\,s^{-1}}$ at $\tau=0$.
  The dots and crosses are the experimental data from Ref. \cite{tsa80} for an empty vessel and one filled with helium II at $T=1.57\,{\rm K}$.
  In both cases, experimental error bars are smaller than the plotted symbols.
  The heavy solid curve is the theoretical solution presented in \S\ref{tsasec2}.
  (Left panel.)  Superfluid coefficients taken from Table 1.
  (Right panel.) As for left panel, but with $B=1.22\times10^{-2}$ and $B'=2.31\times10^{-3}$.}
\label{tsafig1}
\end{figure}

In the left panel of Figure 1, we plot the spin-down recovery of a spherical container of radius $a=1.7\,{\rm cm}$, filled with helium II at $1.57\,{\rm K}$, after an impulsive increase in angular velocity from $\Omega_s=3.0\,{\rm rad\,s^{-1}}$ to $\Omega_s+\delta\Omega=4.2\,{\rm rad\,s^{-1}}$.
The dots and crosses represent experimental data for the empty vessel and helium-filled vessel respectively, corresponding to curves 3 and 2 in Figure 7 of Ref. \cite{tsa80}.
The slope of the spin-down curve for the empty container as a result of friction in the apparatus is $\delta \Omega \Omega_s E^{1/2} T_0=-0.002\,{\rm rad\,s^{-2}}$.
The theoretical spin down for a spherical container calculated from (\ref{tsa203}) is over plotted as a heavy solid curve, using the values of the superfluid coefficients quoted in Table 1.
The crosses and solid curve approach a constant slope $\dot{\Omega}=\delta\Omega \Omega_s E^{1/2} T_0 (1+K)^{-1}$ asymptotically.
Comparing the asymptotic slopes of the crosses and dots gives $K=0.8$.
\begin{table} [h!]
\caption{Superfluid coefficients for helium II at $T=1.57\,{\rm K}$}
\label{tab1}
\begin{tabular}{cccc}
\hline\noalign{\smallskip}
Quantity & Value & Units & Source  \\
\noalign{\smallskip}\hline\noalign{\smallskip}
$\rho_n$ & $21.2 $ & ${\rm kg\,m^{-3}}$ &   Maynard et al. \cite{may76} \\
$\rho$ & $145.2$ & ${\rm kg\,m^{-3}}$ &   Maynard et al. \cite{may76} \\
$B$ & $1.222$ &  &   Barenghi \& Donnelly \cite{bar83} \\
$B'$ & $0.231$ &  &   Barenghi \& Donnelly \cite{bar83} \\
$\eta$ & $1.26\times10^{-6}$ & ${\rm Pa\, s}$ &   Tough et al. \cite{tou63} \\
\noalign{\smallskip}\hline
\end{tabular}
\end{table}

The left panel of Figure 1 demonstrates qualitative agreement between theory and experiment.
The heavy solid curve parallels the crosses asymptotically, but the vertical intercepts extrapolated back from the asymptotic solutions disagree by $0.32\,{\rm rad\,s^{-2}}$.
It is temping to speculate that the discrepancy arises because we are adopting inappropriate values of the hard-to-measure transport coefficients $\eta$, $B$ and $B'$.
For example, $B$ and $B'$ decrease by several orders of magnitude when the Donnelly-Glaberson instability \cite{gla74} is excited by a counterflow and a vortex tangle forms \cite{tsu04}.

To test this idea, we reduce $B$ and $B'$ $10^{-2}$-fold and re-plot the data and theory.
The agreement with experiment worsens: the heavy curve now takes longer to reach its constant-$\dot{\Omega}$ asymptote, and the vertical intercept of the back-extrapolated asymptote is $0.1\,{\rm rad\,s^{-1}}$ lower than in the left panel.
Exhaustive trials reveal that the experimental data cannot be matched by changing the superfluid coefficients.
% 
% The large discrepancy between heavy dots and curves implies that there is much more dissipation predicted by the theory than observed in the experiment.
% The solution (\ref{tsa203}) is self-consistent and conserves angular momentum in the absence of an external torque, as discussed in Ref. \cite{van11}.
% When a viscous torque is included there is dissipation, the amount of which varies depending on the shape of the spin-down curve, and hence the geometry and superfluid coefficients.
% However, this can only account for the difference between the blue-dashed and blue curves; it cannot explain the difference between the blue and blue-dotted curves.

Another explanation for the discrepancy is that the initial conditions are incorrect. 
An important clue that this is in fact the case follows from close inspection of the experimental data.
When the angular impulse removed by the frictional torque is subtracted, the final angular momentum of the fluid and container is much less than the initial angular momentum, given $\Omega_s=3.0\,{\rm rad\,s^{-1}}$.
It is reported in Ref. \cite{tsa80} that $\Omega$ rises to its maximum at $t \sim 10\,{\rm s}$, whereas the Ekman time is $\sim 40\,{\rm s}$.
Hence the fluid partially spins up before the container is released, partially invalidating the impulsive initial conditions assumed in the theory in \S\ref{tsasec2}.
Because the frictional torque is known, the corrected initial angular velocity of the superfluid can be calculated from angular momentum conservation
\footnote{The `initial' (at $\tau=0$ when the container is released after it is accelerated) angular momentum of the vessel (known) plus fluid (unknown) equals the `final' (at any instant $t_f$ in the asymptotic regime) angular momentum of the vessel [known from $\Omega(t_f)$] plus fluid [known in terms of $\Omega(f_t)$ from the theory in \S\ref{tsasec2} and \cite{van11b} in the asymptotic regime] plus the angular impulse taken away by the friction ($T_0 t_f$).
As the angular impulse is given self-consistently and uniquely by the theory, subtracting it to find the initial angular momentum of the fluid is equivalent to adjusting the single unknown $\Omega_s$ until theory and data agree.}.

\begin{figure} [h!]
  \includegraphics[width=0.45\textwidth]{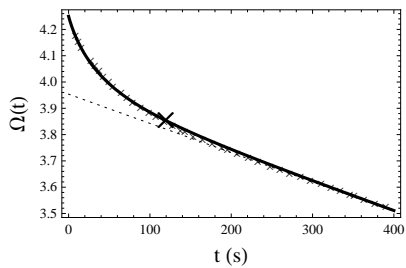} 
\caption{Angular velocity of a spherical container filled with helium II (at $T=1.57\,{\rm K}$) as a function of time, with initial conditions ($\Omega_s=3.68\,{\rm rad\,s^{-1}}$ and $\Omega_s+\delta\Omega=4.25\,{\rm rad\,s^{-1}}$) adjusted to account for the incomplete pre-release spin up.
 The black cross marks the point on the curve where the spin-up time quoted in Ref. \cite{tsa80} is read off.
 The theoretical solution asymptotes to the dotted line.}
\label{tsafig2}
\end{figure}

In Figure 2, we re-plot the solid black curve in Figure 1 with modified initial conditions for the superfluid.
We find that values of $\Omega_s=3.68\,{\rm rad\,s^{-1}}$ and $\Omega_s+\delta\Omega =4.25\,{\rm rad\,s^{-1}}$ give near perfect agreement (error $\approx0.5\%$) between theory and experiment.
In this situation, the superfluid and container rotate at $3.0\,{\rm rad\,s^{-1}}$ before spinning up.
At time $\tau=0$, when the container is released after being accelerated, its angular velocity is $4.25\,{\rm rad\,s^{-1}}$, while the angular velocity of the superfluid has increased to $3.68\,{\rm rad\,s^{-1}}$.
% A slight change of the initial velocity of the container is also required to fit the data, where $\Omega_s+\delta\Omega=4.25\,{\rm rad\,s^{-1}}$.
The close agreement obtained between theory and experiment in the left panel of Figure 2 is strong evidence in support of the hypothesis that helium II spins up via laminar Ekman pumping.
When the viscous component density fraction is small (e.g. at $T=1.57\,{\rm K}$), the spin-up flow is dominated by Ekman-like circulation in the inviscid component \cite{rei93}.
The result also suggests that vortex pinning and tension (which are neglected in our theoretical analysis) are not needed to reproduce the experimental data in Ref. \cite{tsa80}.

\begin{figure} [h!]
  \includegraphics[width=0.45\textwidth]{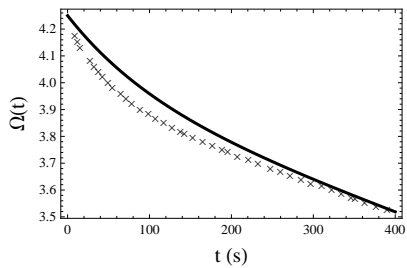} 
\caption{Same as Figure 2, but for a cylinder with $h=a=1.7\,{\rm cm}$ and $\Omega_s=3.8\,{\rm rad\,s^{-1}}$.
The agreement between theory and data is poorer than for a sphere.}
\label{tsafig3}
\end{figure}

We note in passing that the shape of the vessel must be treated correctly, if the theory is to reproduce the data.
Figure 3 displays the spin down of a cylinder filled with helium II at $1.57\,{\rm K}$, which is identical to the apparatus modeled in Figure 2 (with half-height and cylindrical radius $h=a=1.7\,{\rm cm}$) except for its shape.
Even after correcting the initial conditions ($\Omega_s=3.8\,{\rm rad\,s^{-1}}$ and $\Omega_s+\delta\Omega =4.25\,{\rm rad\,s^{-1}}$),
the theory agrees poorly (maximum error $\approx3\%$) with the experimental data.
Spheres and cylinders produce characteristically different spin-down curves, because the torque varies with latitude in a sphere \cite{van11b}.

\subsection{Superfluid spin-up time}

The measured spin-up times $t_0$ for experiments at six different temperatures (with all other experimental parameters identical to Figure 2) are plotted as points with error bars in Figure 4.
The data are taken from curve 2 of Figure 9 in Ref. \cite{tsa80}.
The definition of $t_0$ adopted in Ref. \cite{tsa80} refers to the point at which the spin-down curve reaches its asymptotic regime, drawn as a dotted line in Figure 2.
Clearly, there is scope for ambiguity in identifying exactly when this occurs.
In Ref. \cite{tsa80} and Figure 4, the value of $t_0=119.65\,{\rm s}$ quoted at $T=1.57\,{\rm K}$ corresponds to the point marked with a cross in Figure 2, where the difference between $\Omega(t)$ and the long-term linear spin down drops to $11\%$ of its original value.
To remain consistent, we calibrate against this point and adopt the foregoing definition of $t_0$ throughout the rest of this paper.
\begin{figure} [h!]
  \includegraphics[width=0.45\textwidth]{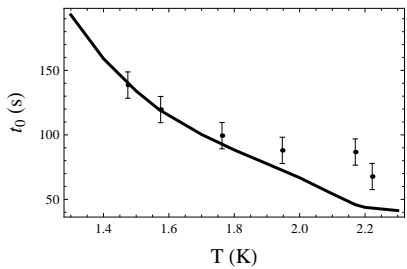}
\caption{Spin-up time $t_0$ (in ${\rm s}$) versus temperature, $T$ (in ${\rm K}$).  Experimental measurements from Ref. \cite{tsa80} are plotted as black points with error bars.
The spin-up time calculated from (\ref{tsa203}) and calibrated against the cross in Figure 2 is plotted as a heavy black curve.}
\label{tsafig4}
\end{figure}

In Figure 4, the theoretical spin-up time [calculated from (\ref{tsa203})] is plotted versus temperature as a heavy black curve.
Good agreement with the data is obtained for temperatures between $1.46\,{\rm K}< T <1.8\,{\rm K}$, where the inviscid component density fraction is less than $0.3$.
Above $2.0 \,{\rm K}$, the viscous component dominates and the agreement worsens.
Turbulent drag may be responsible.
In a classical fluid, the Ekman layer becomes unstable when the Ekman number drops below the critical value \cite{gre68,rei93}
\begin{equation}
 E=\epsilon^2 \left( 56.3+58.4 \epsilon \right)^{-2} \,. \label{tsa301}
\end{equation}
When the flow becomes turbulent, the drag is reduced, lengthening $t_0$.
Above the critical temperature, where the flow is classical, the condition (\ref{tsa301}) applies directly.
Below the critical temperature, transition to turbulence in superfluids is not well understood, although it typically occurs at Ekman numbers similar to (or even higher than) the classical critical value \cite{cou88,she08}.
Reisenegger \cite{rei93} suggested a critical value of $E\approx2.5\times10^{-3}\epsilon^2$, reasoning that the vortex lines should be approximately parallel to the rotation axis in the laminar regime, although it should be noted that this is never strictly allowed by the boundary conditions in a sphere (where the vortex lines emerge perpendicular to the surface).
Under the experimental conditions in Figures 2 and 4, the above results imply that the flow should be turbulent for $T>1.6\,{\rm K}$.

For the spin down of smooth-walled spherical containers, Tsakadze and Tsakadze \cite{tsa80} derived the empirical relation
\begin{equation}
 \Omega_s t_0=A\left(\frac{4 m a^2 \Omega_s}{\hbar}\right)^{\gamma} \left(\frac{\rho_n}{\rho}\right)^{-\alpha}{\rm ln}\left(1+c \delta\Omega\right)\,,\label{tsa303}
\end{equation}
where $m$ is the mass of a helium atom, and $\hbar$ is Planck's constant divided by $2\pi$.  
The remaining constants are fitted by the method of least squares, yielding $A=1.0\pm0.1$, $\gamma=0.40\pm0.05$, $\alpha=0.25\pm0.01$ and $c=5\pm0.2\,{\rm rad^{-1}\,s}$.
In the analytic theory presented in \S\ref{tsasec2} and Ref. \cite{van11a,van11b}, $t_0$ scales as the Ekman time, implying $\Omega_s t_0 \sim (a^2\Omega_s)^{1/2}$, in close agreement with (\ref{tsa303}).
For linear spin-up, $t_0$ does not depend on $\delta\Omega$.
However, in deriving (\ref{tsa303}), flows with $\epsilon$ as high as $0.3$ were studied, where the linearity assumption is questionable, the flow may be turbulent, and $t_0$ does depend on $\delta\Omega$.

The temperature dependence in (\ref{tsa303}) is contained in $\rho_n/\rho$.
\begin{figure} [h!]
  \includegraphics[width=0.45\textwidth]{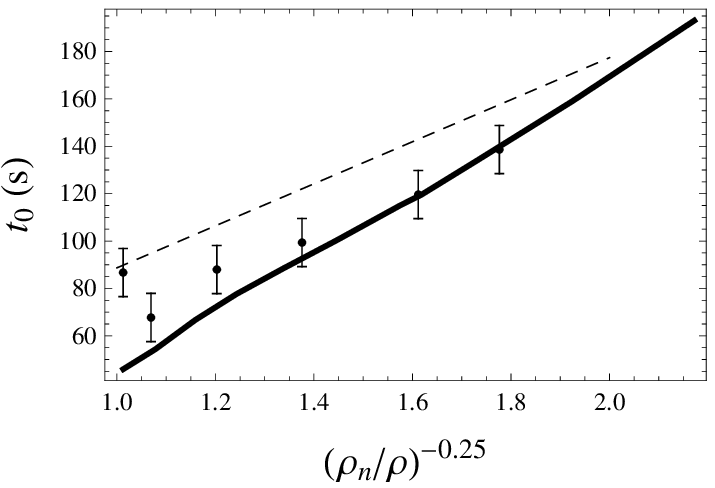}
  \includegraphics[width=0.45\textwidth]{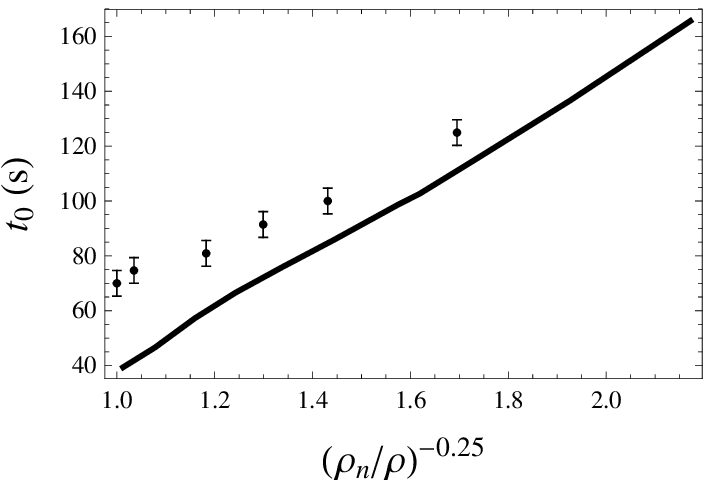}
\caption{Spin-up time $t_0$ (in ${\rm s}$) versus the reciprocal of the fourth root of the density fraction of the viscous component, $\left(\rho_n/\rho\right)^{-0.25}$. 
Left panel: $\Omega_s=3.0\,{\rm rad\,s^{-1}}$, $\Omega_s+\delta\Omega=4.2\,{\rm rad\,s^{-1}}$ (as in Figure 3). 
Right panel: $\Omega_s=5.0\,{\rm rad\,s^{-1}}$, $\Omega_s+\delta\Omega=6.5\,{\rm rad\,s^{-1}}$.
The dashed curves represent the empirical relation (\ref{tsa303}).  Solid curves are the theoretical results calculated from (\ref{tsa203}).  }
\label{tsafig5}
\end{figure}
In the left panel of Figure 5, values from the horizontal axis in Figure 4 are converted from $T$ to $(\rho_n/\rho)^{-0.25}$, using generally accepted values of $\rho(T)$ and $\rho_n(T)$, e.g. contained in Ref. \cite{may76}.
The experimental data are plotted as points with error bars, the theoretical curve [calculated from (\ref{tsa203})] is plotted as a heavy black curve,
and the empirical relation (\ref{tsa303}) is over plotted as a dashed curve.
Note that temperature now increases from right to left.
The critical value $E\approx2.5\times10^{-3}\epsilon^2$ occurs at $(\rho_n/\rho)^{-0.25}=1.58$, below which turbulence is expected.
Equation (\ref{tsa203}) describes the data better than equation (\ref{tsa303}) for these parameters, suggesting that (\ref{tsa303}) is unreliable when extrapolating to parameters outside those for which it is derived.

In the right panel of Figure 5, we compare the theoretical predictions from (\ref{tsa203}) with the specific experiments used to fit $A$, $\gamma$, $\alpha$ and $c$ in (\ref{tsa303}).
A sphere of radius $a=1.7\,{\rm cm}$ is spun up from $\Omega_s=5.0\,{\rm rad\,s^{-1}}$ to $\Omega_s+\delta\Omega=6.5\,{\rm rad\,s^{-1}}$.
The experimental data from Figure 10 in Ref. \cite{tsa80} are plotted as points with error bars.
In this case, the theory departs noticeably from the data.
However, for these parameters the threshold $E\approx2.5\times10^{-3}\epsilon^2$ is achieved at $T=1.25\,K$ and hence $(\rho_n/\rho)^{-0.25}=2.32$.
This suggests that all the data points in the right panel of Figure 4, and therefore $A$, $\gamma$, $\alpha$ and $c$ in the empirical relation (\ref{tsa303}), apply to the turbulent regime.
Unfortunately, because there are no measurements in the laminar regime [$(\rho_n/\rho)^{-0.25}>2.32$], the theory cannot be conclusively tested with these data.

We close by remarking on the radius of the vessel used in the experiments.
In Ref. \cite{tsa80}, a value of $3.4\,{\rm cm}$ is quoted as both the radius and the diameter \cite{rei93}.
In this paper we assume $a=1.7\,{\rm cm}$ everywhere.
The larger value of $a=3.4\,{\rm cm}$ leads to double the spin-up time, making agreement between theory and experiment very difficult to obtain.
Under this interpretation, the radius quoted in equation (6) of Ref. \cite{tsa80} refers instead to the diameter.
An appropriate correction is made in equation (\ref{tsa303}) of this paper.

\section{Conclusions}
\label{tsasec4}

A self-consistent analytic solution of the HVBK equations for the spin up of helium II enclosed in a vessel of arbitrary geometry, including the viscous feedback torque, is compared with experimental data from the classic experiments conducted by Tsakadze and Tsakadze \cite{tsa80}.
Excellent agreement between the theoretical and observed spin-down of the container is obtained at $T=1.57\,{\rm K}$.
The agreement supports the hypothesis that laminar Ekman pumping occurs, which vortex tension is negligible and the container walls are smooth enough to avoid much pinning.
Because the container initially spins up over $\sim10\,{\rm s}$, a fair fraction of the Ekman time ($\sim 40\,{\rm s}$), the acceleration event is not truly impulsive, and the theoretical initial conditions must be corrected by angular momentum conservation arguments \cite{van11b}.

As one approaches the superfluid transition temperature, the theoretical and observed spin-down times diverge.
It is suggested that this discrepancy is caused by the onset of turbulence as $\eta$ and $\rho_n/\rho$ increase, consistent with a critical Ekman number $E\approx2.5\times10^{-3}\epsilon^2$ proposed in Ref. \cite{rei93}.
Several observable differences are present if the flow is turbulent: 
(1) the spin-up time depends on $\delta\Omega$, and
(2) the sphere and cylinder have similar spin-down curves, because latitudinal variation of the hydrodynamic torque in Ekman pumping is wiped out to some extent.
Further experiments, conducted specifically in the range $\epsilon<E^{1/2}\ll1$ are required to test the theory fully.
The coupled spin up of a superfluid and its container is a vital component in understanding the nature of vortex pinning and nucleation in terrestrial settings and the recovery process following rotational glitches in neutron stars.

\begin{acknowledgements}
CAVE acknowledges the financial support of an Australian Postgraduate Award.
Carlos Peralta is thanked for useful discussions regarding the helium II experiments.
\end{acknowledgements}

\bibliographystyle{spmpsci}      % basic style, author-year citations
\bibliography{tsakadze}   % name your BibTeX data base

%\begin{thebibliography}{}
% etc
%\end{thebibliography}

\end{document}